\documentclass{eds2019}
\usepackage{amssymb}

\def\be{\begin{equation}}
\def\ee{\end{equation}}
\def\bea{\begin{eqnarray}}
\def\eea{\end{eqnarray}}

\def\sqs{\sqrt{s}}
\def\sqsn{\sqrt{s_\mathrm{NN}}}
\def\Raa{R_\mathrm{AA}}

\def\pT{p_\mathrm{T}}

\newcommand{\Photo}{}

\begin{document}
\vspace*{4cm}
\title{OVERVIEW OF RECENT ALICE RESULTS}

\author{R\'obert V\'ertesi (for the ALICE Collaboration)}

\address{Wigner Research Centre for Physics, Centre of Excellence of the Hungarian Academy of Sciences\\1121 Budapest, Konkoly-Thege Mikl\'os \'ut 29-33, Hungary}

\maketitle\abstracts{
A selection of recent ALICE results from pp, p--Pb, Pb--Pb and Xe--Xe collisions taken during the Run 1 and Run 2 phases are summarized in this contribution, and the prospects for the upcoming Run 3 phase are outlined.}

\section{Introduction}

It was long anticipated that at sufficiently high temperatures and energy densities the strongly interacting matter would no longer be confined into hadrons.\cite{Hagedorn:1965st,Bondorf:1978kz} Such deconfined matter, the Quark--Gluon Plasma (QGP) is assumed to have filled the early universe after the first microseconds.
In the past decades, large experiments at RHIC and the LHC have shown that the QGP exhibits strong collective behavior, similar to an extremely hot and almost perfect fluid.\cite{Adcox:2004mh,Aamodt:2010pa} 
Data from the Run 2 phase of the LHC allowed for precision measurements aimed at a detailed understanding of QGP properties.
ALICE is a dedicated heavy-ion experiment at the CERN LHC accelerator with excellent identification capabilities in collisions with high particle multiplicities in the final state.\cite{Abelev:2014ffa} This contribution summarizes some of the most intriguing results.

\section{Production of identified particles}

\begin{figure}[h]
	\center
	\includegraphics[width=0.6\columnwidth]{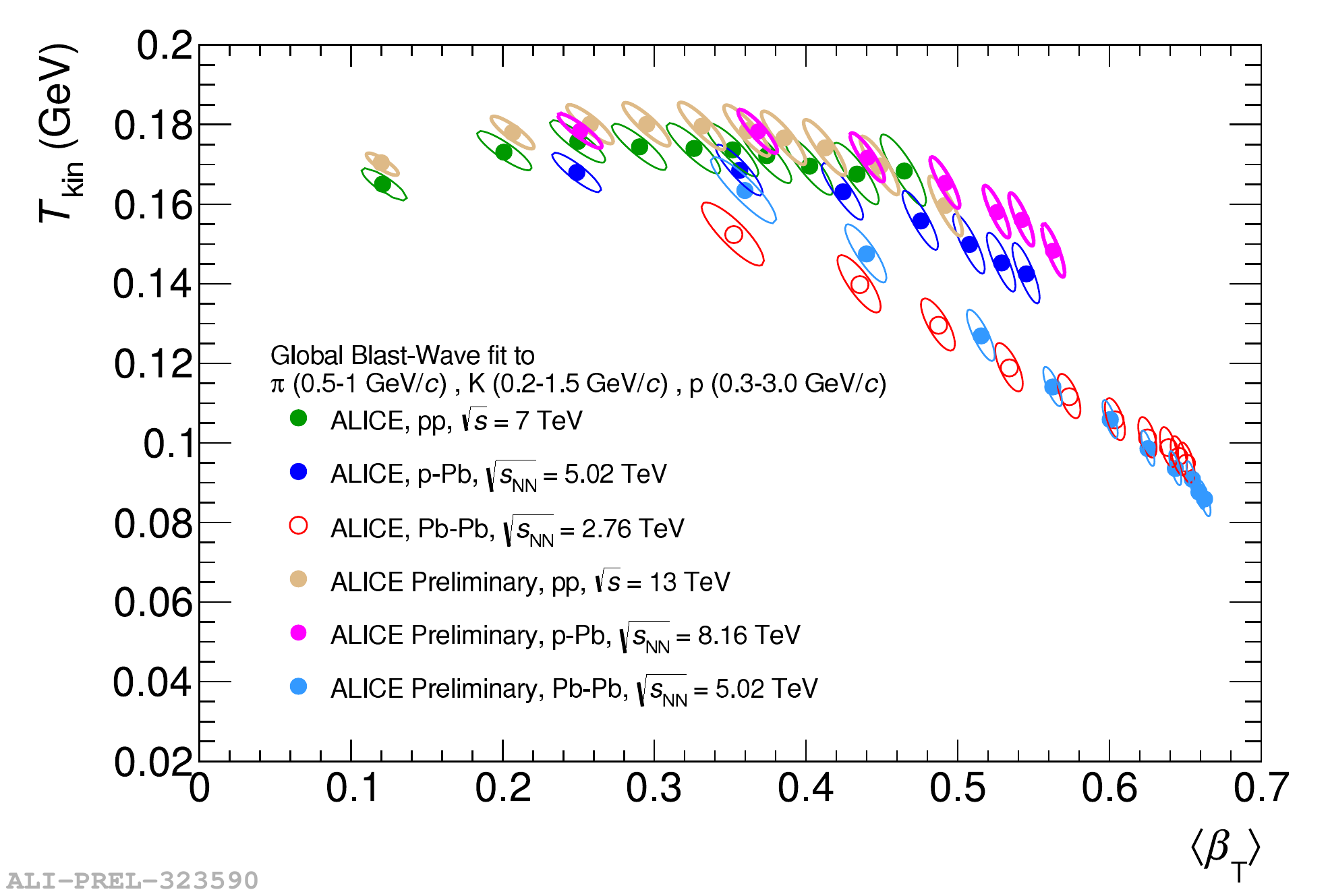}%
	\includegraphics[width=0.4\columnwidth]{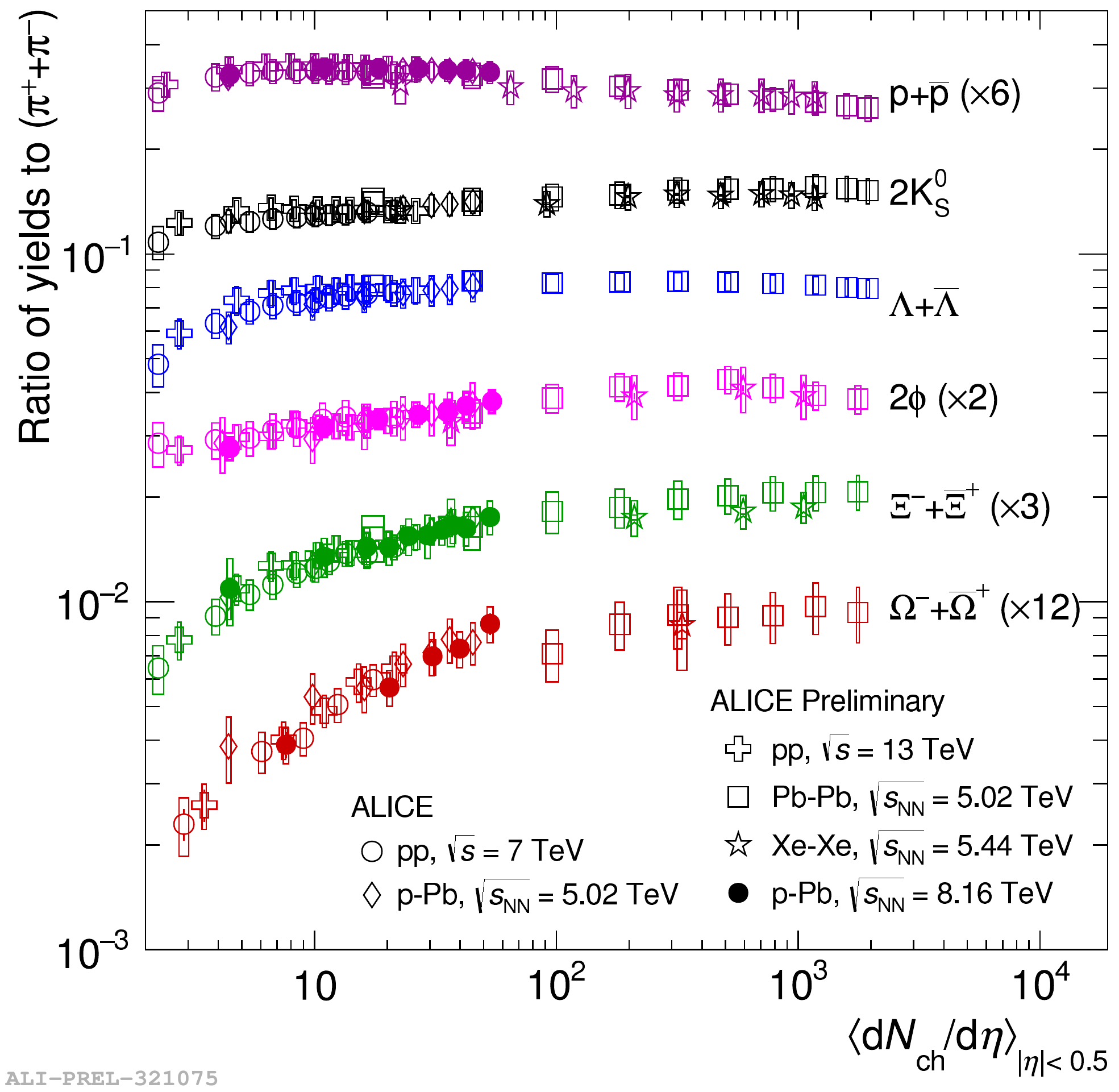}%
	\caption{\label{fig:ToPionRatios}%
		{\it Left:} $T_{\rm kin}$ and $\beta_{\rm T}$ parameters from blast-wave fits in different colliding systems and collision energies. {\it Right:} p, $\mathrm{K}^0_{\rm S}$, $\Lambda+\bar{\Lambda}$, $\Xi^-+\bar{\Xi}^+$, $\Omega^-+\bar{\Omega}^+$ and $\phi$ to $\pi^\pm$ ratios in function of event multiplicity in pp collisions at $\sqs$ = 7 and 13 TeV, p-Pb collisions at $\sqsn$ = 5.02 and 8.16 TeV, PbPb collisions at $\sqsn$ = 5.02 TeV and Xe-Xe collisions at $\sqsn$ = 5.44 TeV.}
\end{figure}

ALICE carried out a broad set of high-precision measurements of identified particles at several collision energies and in different colliding systems.\cite{Acharya:2018qsh,Acharya:2018eaq,Aamodt:2010my} The mass-dependent hardening of light-particle spectra with increasing multiplicity suggests that spectral slopes are determined by a statistical freezeout temperature that is modified by the radial expansion of the freezeout surface. An oft-used parametrization is the blast-wave model, where particles are produced on an expanding hypersurface.\cite{Schnedermann:1993ws} The spectra can then be determined by the radial expansion velocity $\beta_{\rm T}$ and the kinetic freeze-out temperature $T_{\rm kin}$. 
The results of a simultaneous fit to the spectra of light particles are shown in Fig.~\ref{fig:ToPionRatios} (left) as a function of multiplicity for various collision systems and energies.
Although the trends are similar in all three collision systems, similar values of expansion velocity correspond to smaller freeze-out temperatures in small than in large systems. On the other hand, the dependence on collision energy is weak within a given collision system.

Strangeness enhancement was traditionally considered as a smoking-gun signature of the QGP formation.\cite{Rafelski:1982pu}
Figure~\ref{fig:ToPionRatios} (right) summarizes ALICE measurements of strange and non-strange hadron yields normalized by the yield of pions, across several collision systems and energies as a function of charged-hadron event multiplicity at mid-rapidity. There is a clear sign of enhancement that increases with strangeness content. However, no significant energy and system dependence is present at any given multiplicity, and a universal smooth evolution can be observed with event multiplicity regardless of collision system or energy.
These observations suggest that the production of light and strange particles are driven by the characteristics of the final state. An implication of this is that penetrating probes are required to learn about the onset and the nature of QGP production.

\section{Collective phenomena in small and large systems}

\begin{figure}[h]
	\center
	\includegraphics[width=0.6\columnwidth]{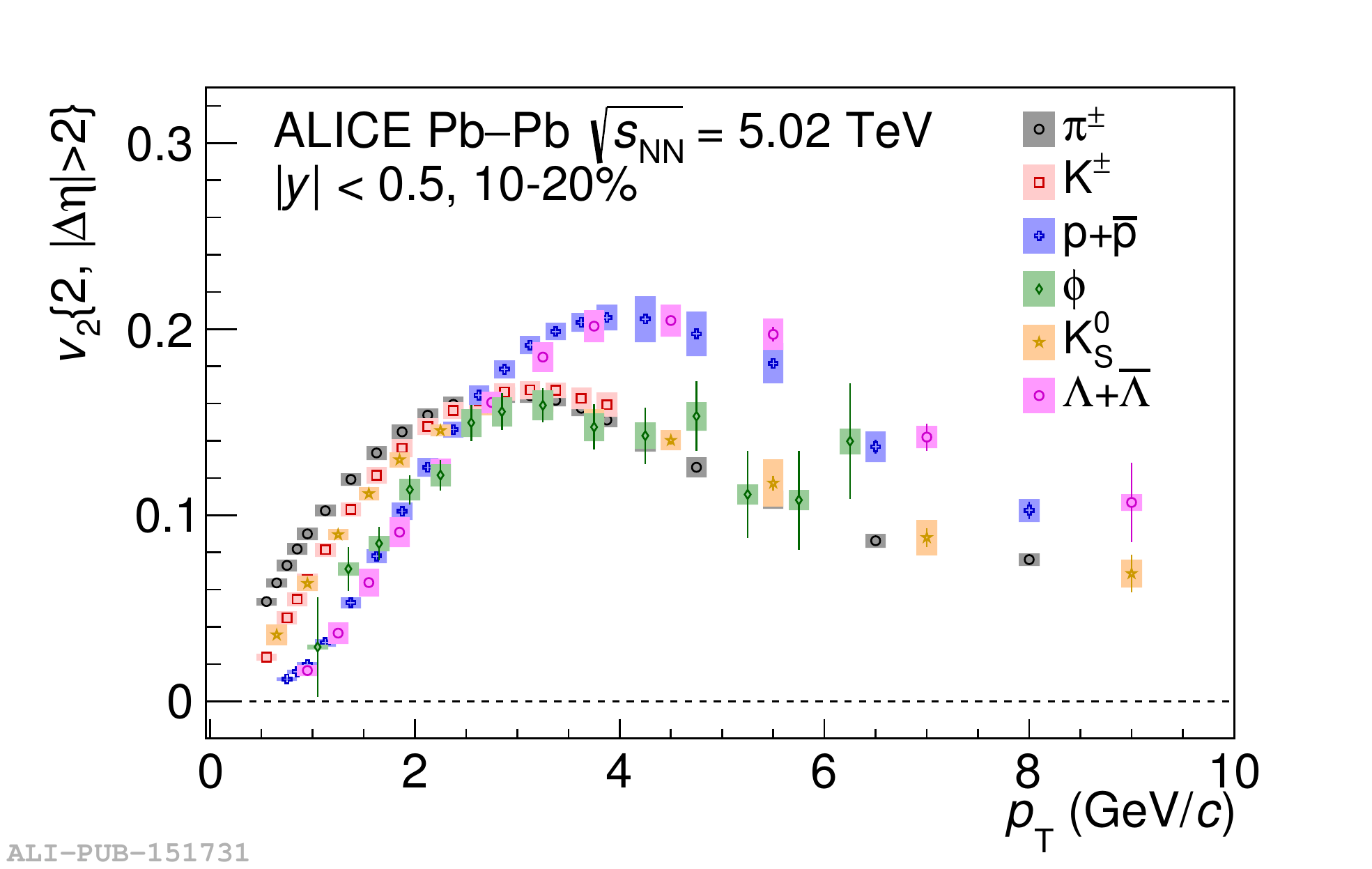}%
	\includegraphics[width=0.4\columnwidth]{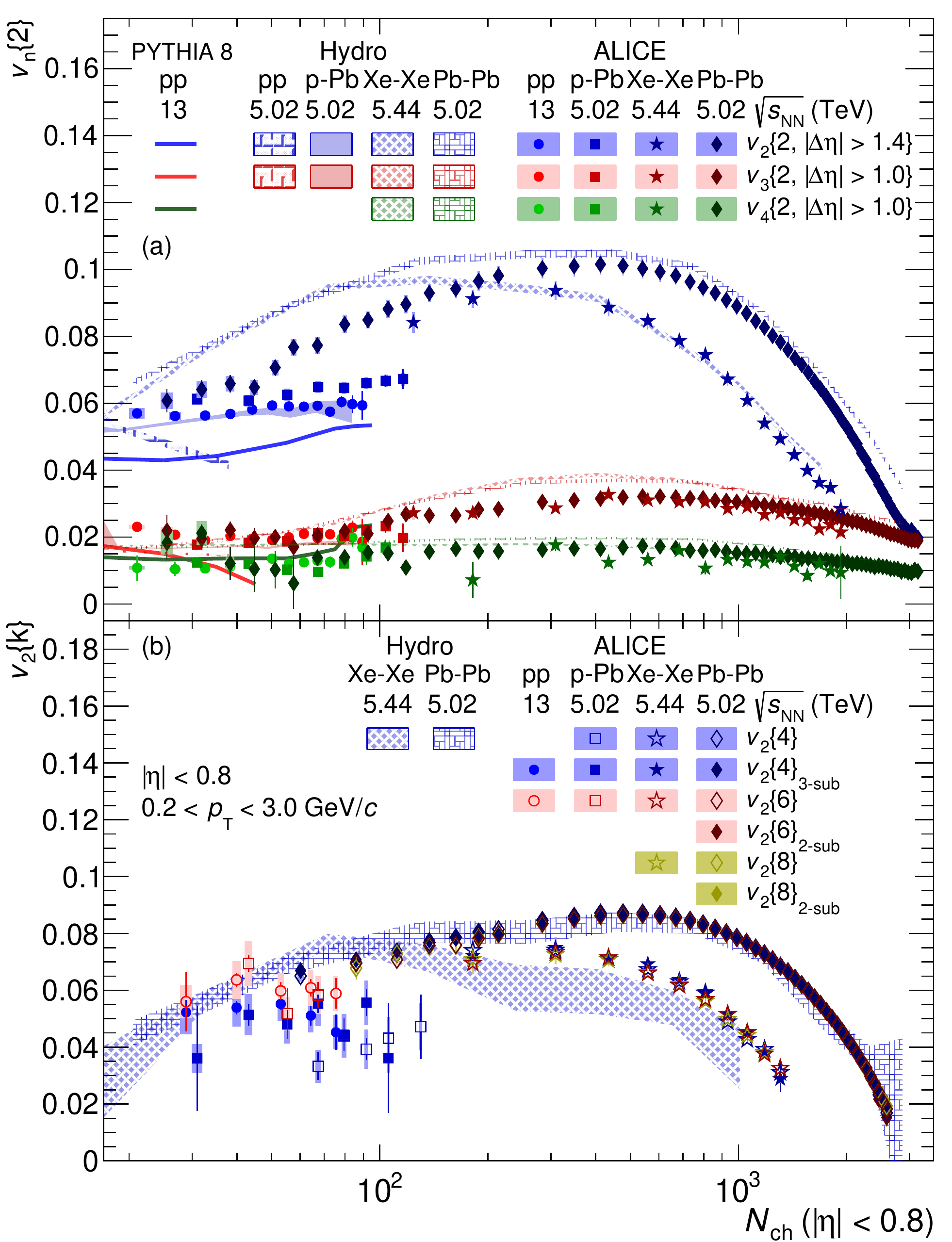}%
	\caption{\label{fig:vnCoeffs}%
		{\it Left:} The $\pT$-differential $v_2$ of $\pi^\pm$, K$^\pm$,  K$^0_s$, p+$\bar{\mathrm{p}}$, $\Lambda$ and $\bar{\Lambda}$ and $\phi$ in 10--20\% centrality Pb--Pb collisions at $\sqsn$ = 5.02 TeV. {\it Top right:} Multiplicity dependence of $v_n$ obtained with two-particle cumulants for pp collisions at $\sqsn$ = 13 TeV, p--Pb and Pb--Pb collisions at $\sqsn$ = 5.02 TeV and Xe--Xe collisions at $\sqsn$ = 5.44 TeV. {\it Bottom right:} Multiplicity dependence of $v_2$ coefficients obtained with multiparticle cumulants.}
\end{figure}

In the picture of the strongly interacting QGP emerging in the era of RHIC, collective phenomena were associated with the production of the QGP in high-energy heavy-ion collisions. The LHC experiments, however, discovered several collective features in smaller pp and pA systems with sufficiently high multiplicity.\cite{Khachatryan:2016txc,Abelev:2012ola}
The azimuthal momentum anisotropy of the final-state particles, also called flow, is often described in a Fourier decomposition.\cite{Voloshin:1994mz} 
While a substantial second Fourier component $v_2$ (``elliptic flow'') has been traditionally associated with the collective motion of the final state, the presence of higher-order (especially the odd) coefficients highlight the importance of the initial state in the development of azimutal anisotropy.\cite{Takahashi:2009na} In fact, $v_n$ are sensitive to the full evolution of the system from initial conditions through the QGP until the hadronic phase.
Figure~\ref{fig:vnCoeffs} (left) shows the comparison of $v_2$ coefficients for several particle species in semi-central Pb--Pb collisions at $\sqsn=5.02$ TeV.\cite{Acharya:2018zuq} At low $p_\mathrm{T}$ a clear mass ordering of $v_2$ is present. At intermediate $\pT$ (in the range between $2.5 \lesssim\pT\lesssim 6$ GeV/$c$) an approximate constituent-quark-number scaling can be observed, that is, the baryons and the mesons group together, the two groups being distant from each other. Above $\pT\approx 6$ GeV/$c$, however, parton energy loss becomes dominant and the scaling falls apart.
The right panels of Fig.~\ref{fig:vnCoeffs} present the $v_n$ coefficients in pp, p--Pb, Xe--Xe, and Pb--Pb systems.\cite{Acharya:2019vdf} Long-range multiparticle correlations are clearly observed in all systems, and the two-particle, multi-particle and subevent methods yield qualitatively the same results. The slight systematic difference between the two-particle method and the other methods is owed to non-flow contribution (non-collective correlations).
The ordering of $v_2$, $v_3$ and $v_4$ are the same regardless of systems, and there is a quantitative match of the $v_n$ coefficients throughout the systems at low charged-hadron multiplicity ($N_{\rm ch}$). At higher $N_{\rm ch}$ values, however, $v_2$ does not scale with $N_{\rm ch}$, which suggests different initial geometries in small and large systems. Also, neither pQCD nor hydrodynamics-based models\cite{Sjostrand:2014zea,Mantysaari:2017cni} provide a satisfactory description of pp and p--Pb data.

\section{Medium interactions}

Interactions of high-$\pT$ self-generated probes with the hot medium have traditionally been addressed by the measurement of nuclear modification factors, $\Raa$, where the yields of particles or jets in heavy-ion collisions are compared to reference yields in pp collisions, scaled by the average number of binary nucleon--nucleon collisions within a nucleus--nucleus collision.
While the $\Raa$ is sensitive to hadronization and radial flow at low $\pT$, a universal high-$\pT$ suppression has been found among all the light and strange hadrons at RHIC and the LHC,\cite{Adler:2006hu,Adam:2017zbf} which can be associated to parton energy loss in the colored medium.
The high delivered luminosities and the high-precision capabilities of the current experiments have recently opened the possibility for measuring more refined observables such as correlation or jet structure observables, which aim for the study of jet development within the medium. Grooming techniques allow us to understand hard jet substructures while mitigating the effects of soft fragmentation.\cite{Asquith:2018igt} 

\begin{figure}[h]
	\center
	\includegraphics[width=\columnwidth]{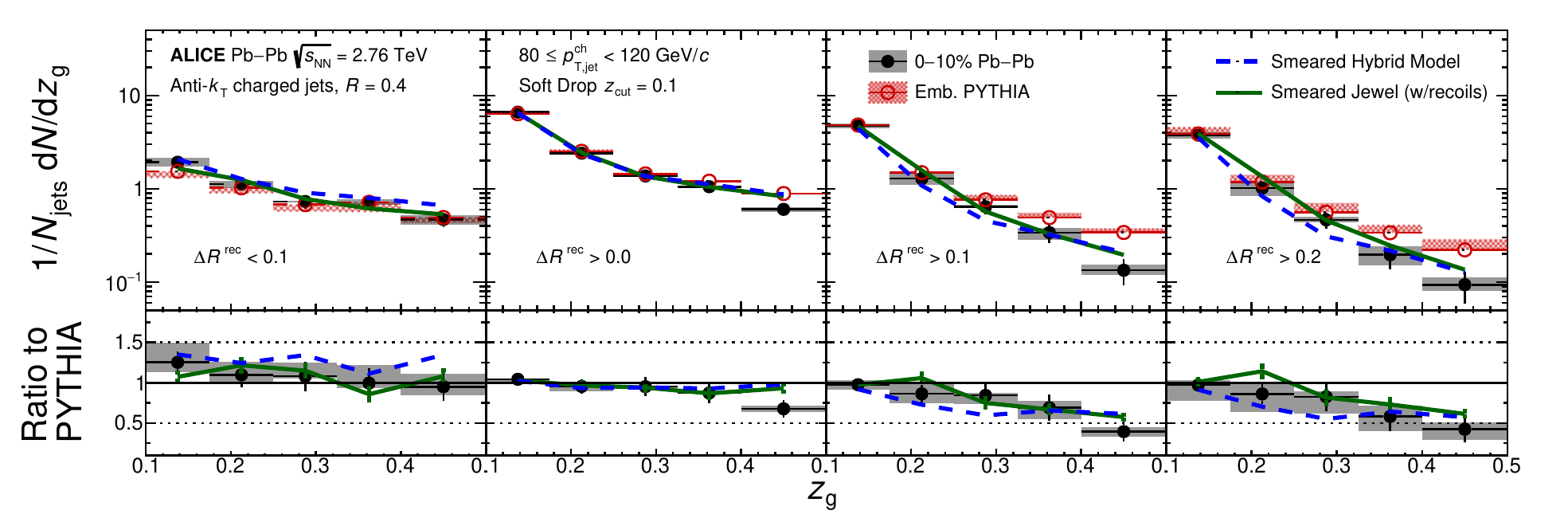}%
	\caption{\label{fig:JetSubstruct}%
		Detector-level Pb--Pb distributions of $z_g$~for $R$=0.4 jets with varying minimum/maximum angular separation of subjets ($\Delta R$) for jets in the charged jet momentum range $80\le \pT\le 120$ GeV/$c$, compared to model calculations.
	}
\end{figure}

ALICE has measured the jet substructure variable $z_g=\frac{min({\pT}_1,{\pT}_2)}{{\pT}_1+{\pT}_2}$,
where ${\pT}_1$ and ${\pT}_2$ are the leading and subleading prongs from the first intra-jet splitting determined using an iterative declustering.\cite{Acharya:2019djg}
Figure~\ref{fig:JetSubstruct} shows $z_g$ distributions in central Pb--Pb collisions at $\sqsn=2.76$ TeV, in four different categories by the opening angle between the two prongs, $\Delta R$.
While embedded PYTHIA pp simulations\cite{Sjostrand:2014zea} describe Pb--Pb data generally well, there is a reduction of small-angle splittings and an enhancement of large-angle splittings in data compared to embedded simulations. Models that include the medium response from jet-medium interactions provide a better agreement with the data.\cite{KunnawalkamElayavalli:2017hxo} This highlights the importance of the interplay between early jet development and the medium.

\section{Direct photons}

\begin{figure}[h]
	\center
	\includegraphics[width=0.4\columnwidth]{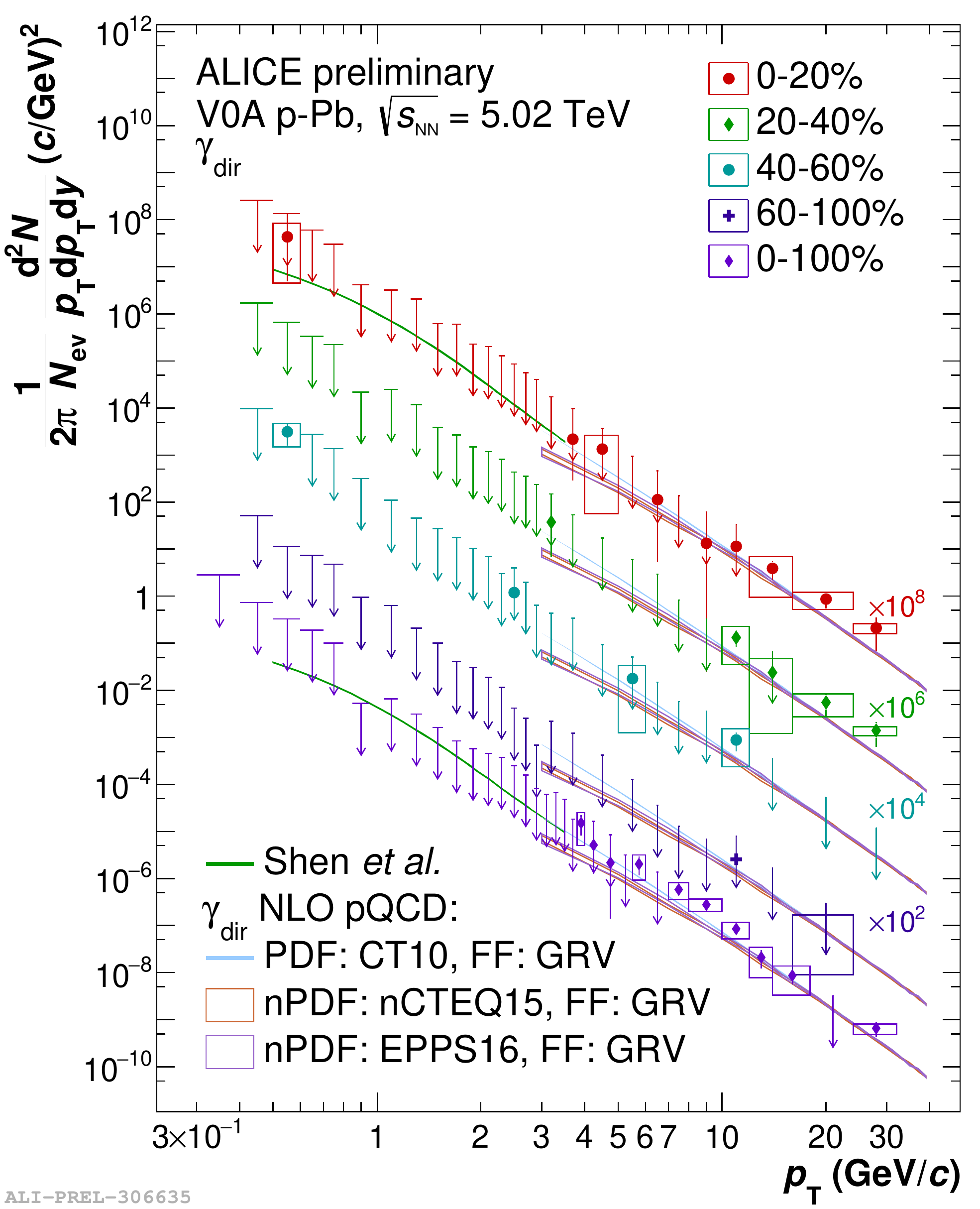}%
	\hspace{10mm}
	\includegraphics[width=0.5\columnwidth]{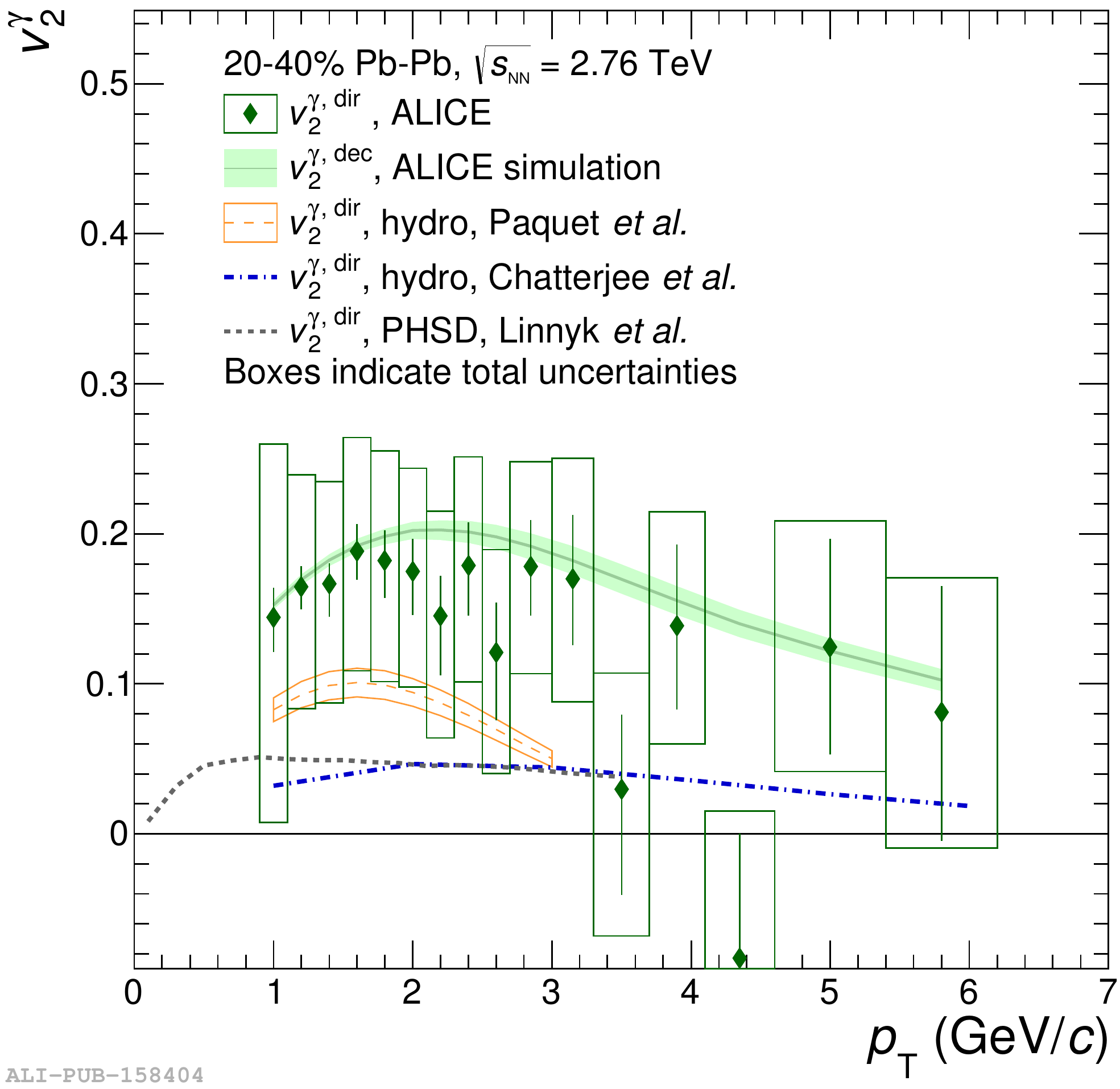}	\caption{\label{fig:DirectPhotons}%
		{\it Left:} Invariant yield of the measured direct photons for several multiplicity bins and the full non-single diffractive sample of p--Pb collisions at $\sqsn=5.02$ TeV, compared to models.
		{\it Right:} Elliptic flow of direct photons compared to the expected flow of decay photons as well as model calculations in the 20--40\% centrality class.%
	}
\end{figure}

The strongly interacting deconfined matter created in high-energy heavy-ion collisions is transparent to electromagnetic particles. Direct photons (photons not coming from hadron decays) are therefore able to bring information from all stages of the reaction including hard scattering, jet radiation, the QGP, as well as the hadron gas. An excess in the low-$\pT$ direct photon spectrum above the yields expected from pp measurements is attributed to the thermal radiation of the hot medium, and implies the presence of the QGP\cite{Adare:2008ab,Adam:2015lda} with an initial temperature between 300 and 600 MeV in central Pb--Pb collisions at $\sqsn=2.76$ TeV. Figure~\ref{fig:DirectPhotons} (left) shows recent ALICE measurements of direct photon yields in p--Pb collisions. No excess is present in the thermal region above pQCD-based models with cold nuclear matter effects, thus corroborating the above interpretation.
Figure~\ref{fig:DirectPhotons} (right) shows the azimuthal anisotropy of direct photons in semi-central Pb--Pb collisions at $\sqsn=2.76$ TeV. The $v_2$ of direct photons is in agreement with that of hadron decay photons. All current models, including those that assume the dominance of late stages in the observed flow,\cite{Shen:2016zpp} predict lower flow for the direct photons. This observation questions our current understanding of the role of thermal photons.

\section{Heavy-flavor mesons and quarkonia}

Heavy-flavor (charm and beauty) quarks are produced almost exclusively in early hard processes. Measurements of heavy-flavor production in small systems collisions can therefore be used as benchmarks of perturbative quantum-chromodynamics (QCD) models. Heavy-flavor particles, especially when compared to light flavor, also provide insight to softer QCD mechanisms like multiple-parton interactions and flavor-dependent fragmentation. Because of their long lifetime, in collisions where there is a nuclear medium, they can be used as self-generated penetrating probes that provide us with means to understand the properties of hot and cold nuclear matter (in nucleus--nucleus and proton--nucleus collisions, respectively). While the high-$\pT$ range mostly brings information about the collisional and radiative energy loss mechanisms in the perturbative regime, measurements at lower $\pT$ can address collective behavior and give insight to coalescence mechanisms between heavy and light flavor.\cite{Andronic:2015wma}

\begin{figure}[h]
	\center
	\includegraphics[width=.5\columnwidth]{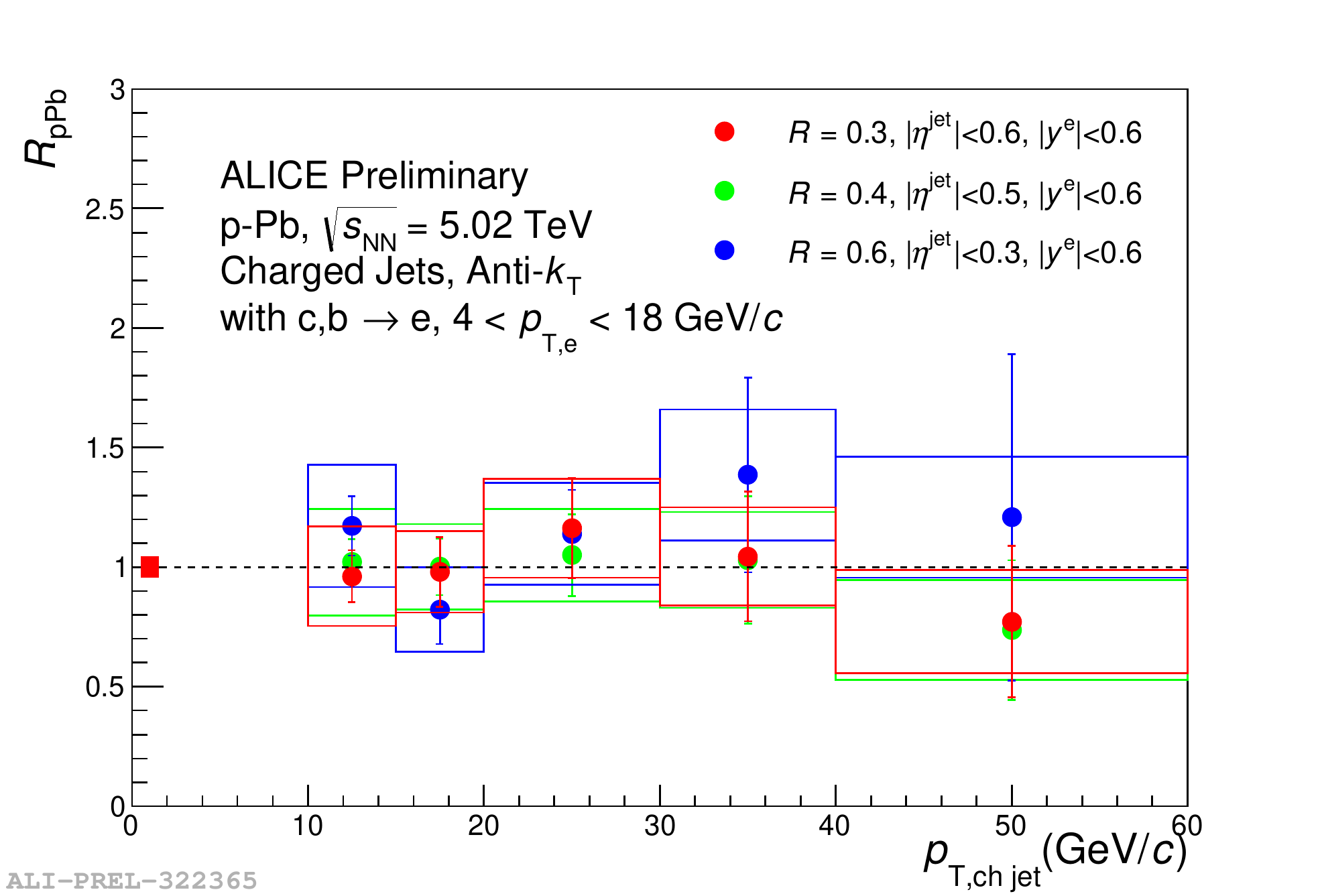}%
	\includegraphics[width=.5\columnwidth]{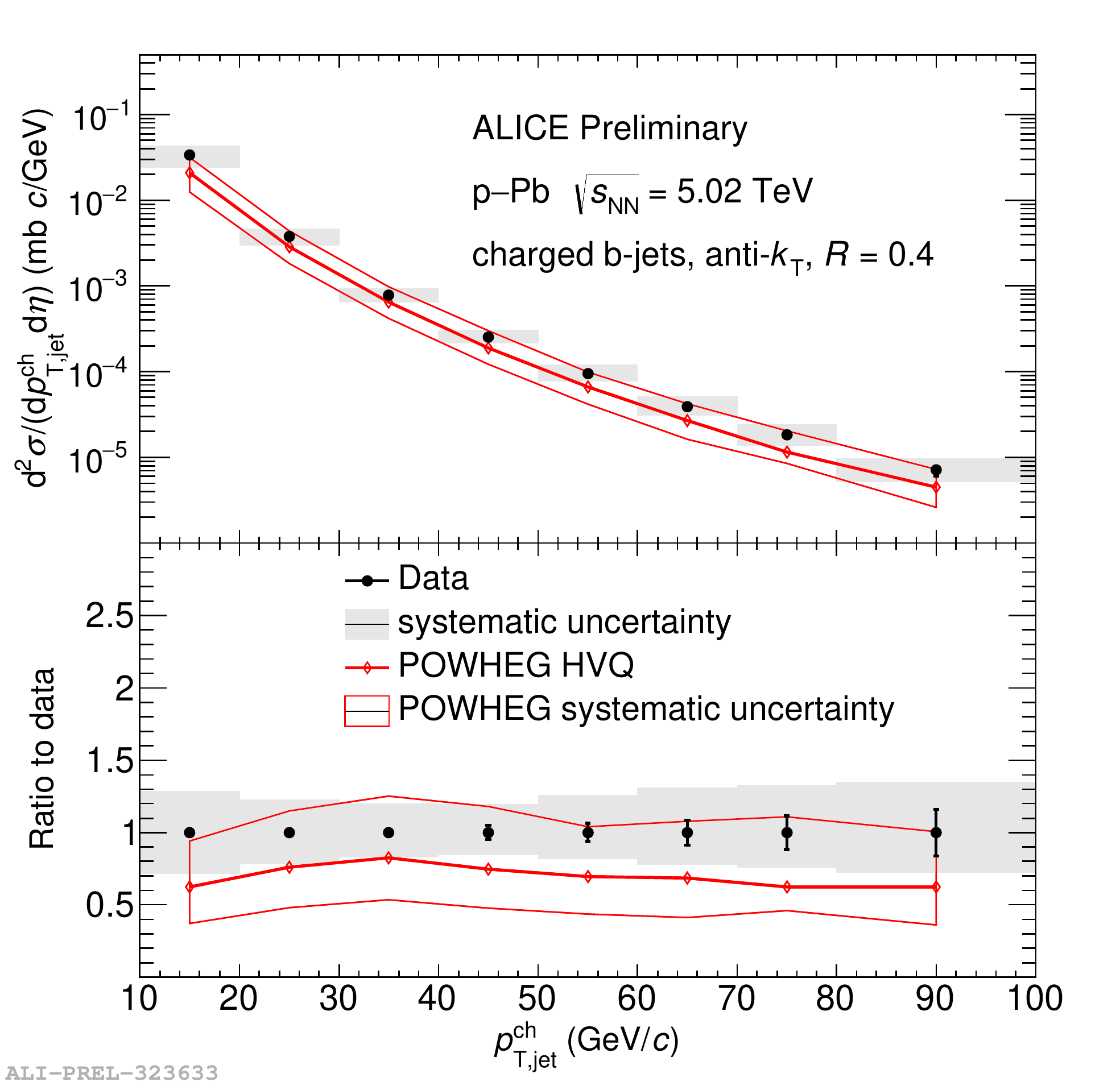}
	\caption{\label{fig:pPbHFjets}%
		{\it Left:} Nuclear modification factor of jets containing a HFE in p--Pb collisions ad $\sqsn=5.02$ TeV, reconstructed with resolution parameters $R = 0.3$, 0.4, and 0.5. {\it Right:} Cross-section of beauty-jets in p--Pb collisions at $\sqsn=5.02$ TeV, reconstructed with the anti-$k_{\rm T}$ algorithm with a resolution parameter $R=0.4$, obtained with secondary vertex tagging. The data are compared to the POWHEG HVQ model scaled by the Pb nuclear mass number, and their ratio is shown on the bottom panel.
	}
\end{figure}

Both the ALICE heavy-flavor electron (HFE) and the D-meson measurements in p--Pb collisions agree with the expectations from pp collisions, suggesting that charm production is not modified substantially by the cold nuclear matter effects at mid-rapidity.\cite{Adam:2015qda,Adam:2016ich} Figure~\ref{fig:pPbHFjets} (left) shows 
measurements on the nuclear modification of jets containing a heavy-flavor electron. Regardless of the choice of the jet resolution parameter, the corresponding $R_{\rm pPb}$ is consistent with unity. 
Figure~\ref{fig:pPbHFjets} (right) shows that the cross-section of jets containing a beauty quark is consistent with POWHEG HVQ\cite{Frixione:2007nu} pQCD-based predictions. Although the uncertainties are rather sizeable, these new results indicate that the production of jets initiated by charm and beauty quarks is not influenced strongly by the presence of cold nuclear matter.

\begin{figure}[h!]
	\center
	\includegraphics[width=0.45\columnwidth]{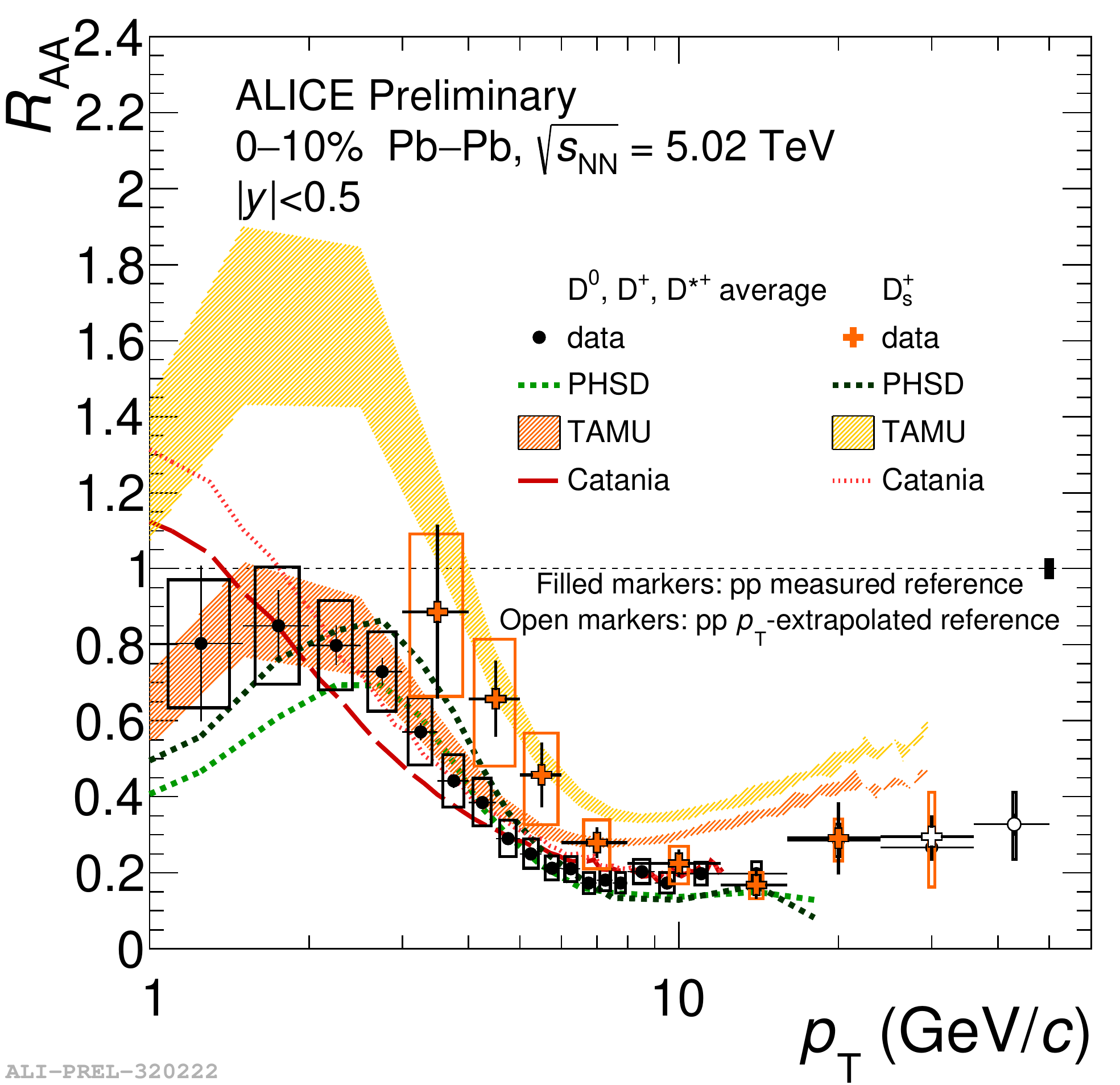}%
	\hspace{0.05\columnwidth}%
	\includegraphics[width=0.45\columnwidth]{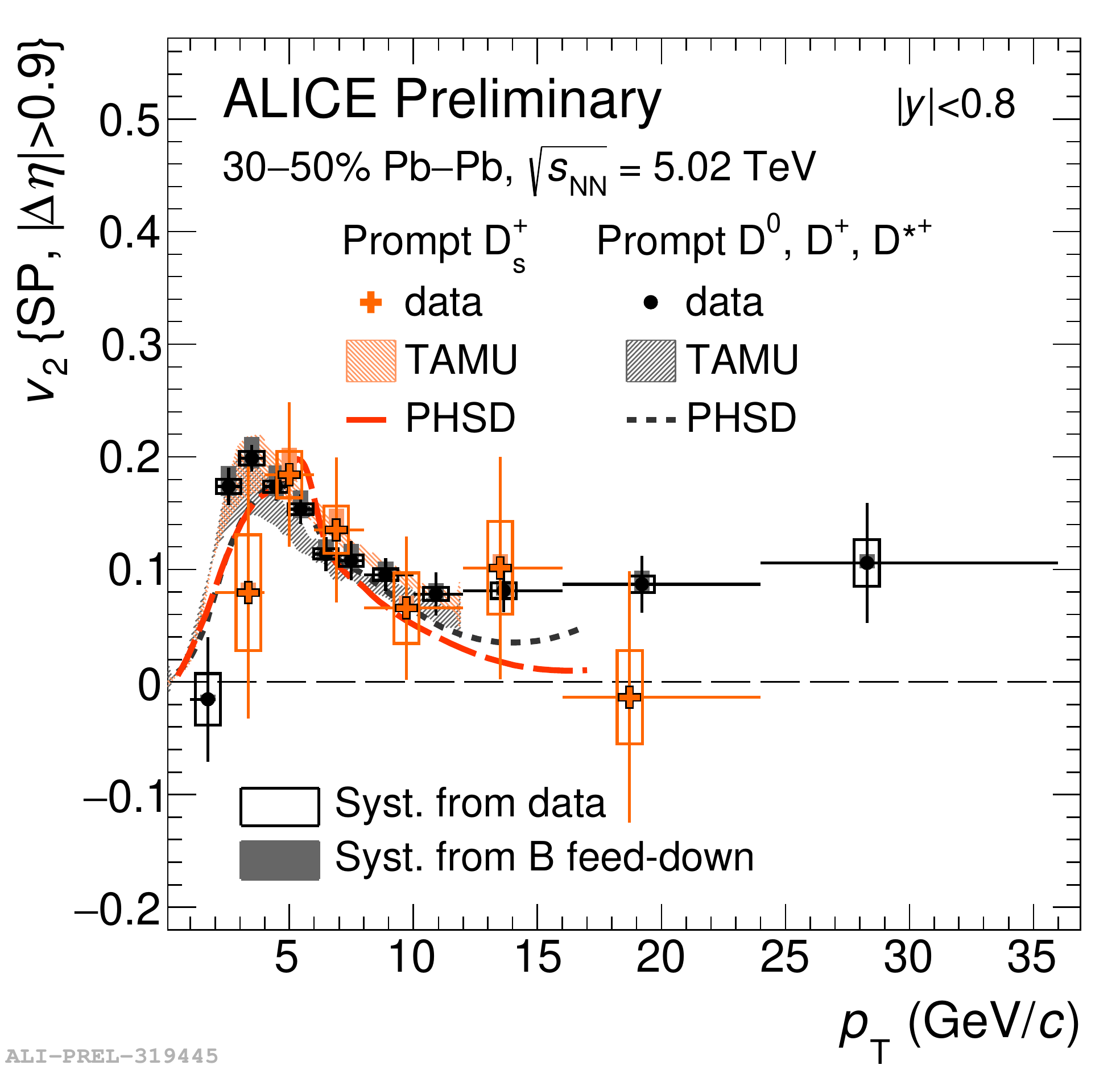}%
	\caption{\label{fig:Dmesons}%
		{\it Left}: Average non-strange D-meson $\Raa$ and prompt ${\rm D}_{\rm s}^+$ $\Raa$ in 30--50\% central Pb--Pb collisions at $\sqsn = 5.02$ TeV, compared with theoretical predictions from transport models. 
		{\it Right}: Prompt ${\rm D}_{\rm s}^+$ and average non-strange D-meson $v_2$ as a function of $\pT$ in the 30--50\% centrality class of Pb--Pb collisions at $\sqsn = 5.02$ TeV, compared to models implementing heavy-quark transport in an hydrodynamically expanding medium.
	}
\end{figure}

Figure~\ref{fig:Dmesons} presents the nuclear modification factor $\Raa$ and the azimuthal anisotrophy parameter $v_2$ of non-strange as well as strange D mesons. At high $\pT$, a substantial suppression can be observed, which is consistent with that of light mesons (not shown). That no mass ordering is present between light and heavy flavors contradicts na\"{i}ve expectations of the ordered energy loss mechanisms, but can be understood by models taking dead cone and color charge fragmentation effects into account.\cite{Djordjevic:2014hka}
Focusing on the low-$\pT$ regime, the D mesons show less suppression than light flavor, and there is an indication of weaker suppression of strange than non-strange D mesons. At the same time, both strange and non-strange D mesons exhibit azimutal anisotropy that is comparable to that of light mesons. This is well-described by models assuming the coalescence of charm with light quarks in an environment of relative strangeness enhancement.\cite{He:2014cla,Song:2015sfa,Plumari:2017ntm}

\begin{figure}[h!]
\begin{minipage}{0.48\columnwidth}
	\center
	\includegraphics[width=.86\columnwidth]{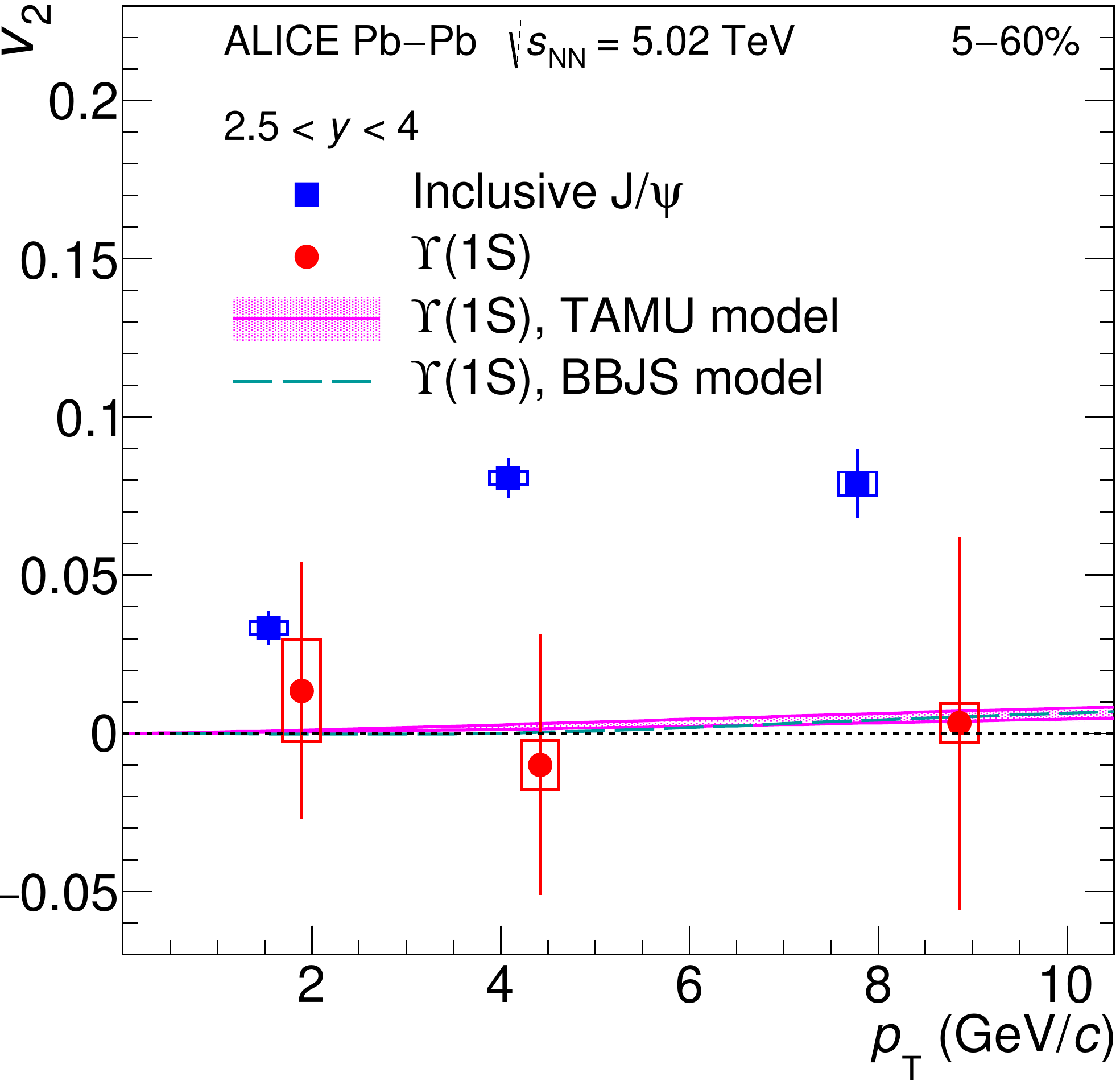}%
	\caption{\label{fig:OniumFlow}%
		Elliptic flow of $\Upsilon(1S)$ mesons in Pb--Pb collisions at $\sqsn = 5.02$ TeV in function of $\pT$, compared to that of $J/\psi$ mesons as well as to model calculations.
	}
\end{minipage}
\hfill
\begin{minipage}{0.5\columnwidth}
	\center
	\vspace{-7mm}
	\includegraphics[width=\columnwidth]{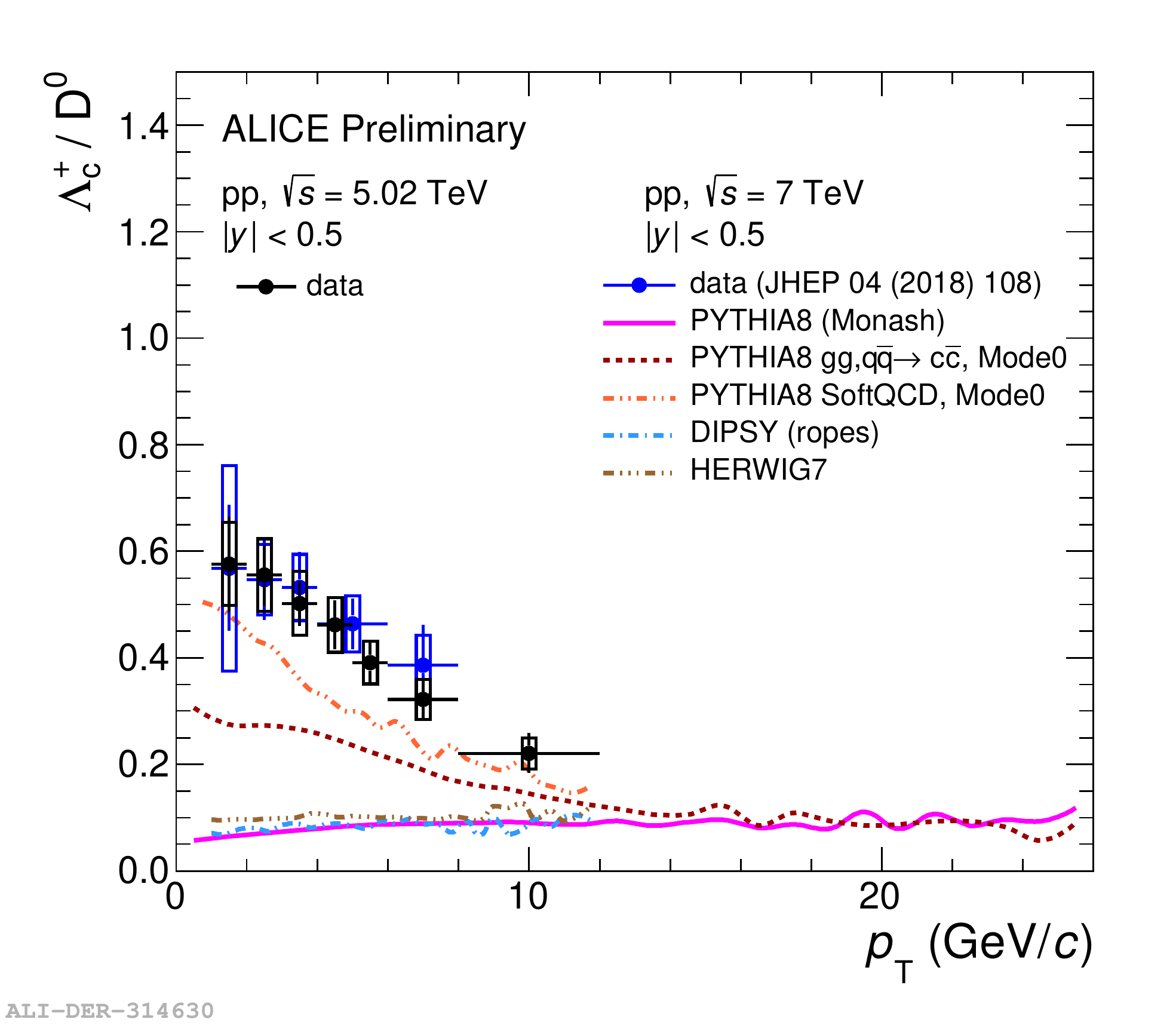}%
	\caption{\label{fig:LambdaC}%
	The $\Lambda^+_{\rm c}/{\rm D}^0$ ratio as a function of $\pT$ measured in pp collisions at $\sqs= 5.02$ and 7 TeV, compared to model calculations.
	}
\end{minipage}
\end{figure}

While open heavy flavor can be used mostly for the tomographic study of the medium, one can address the thermodynamical properties of the QGP by looking at the production of quarkonia (bound states of heavy quarks and their antiquark pairs). 
Sequential suppression of different quarkonium states in a colored medium by the Debye-screening of Q$\bar{\rm Q}$ potential has long been proposed as a sensitive thermometer of the QGP.\cite{Mocsy:2007jz}
The $\Upsilon$ bottomonium states are found to follow the predicted sequential suppression pattern in both RHIC and LHC heavy-ion collisions.\cite{Adamczyk:2016dzv,Chatrchyan:2012lxa,Acharya:2018mni}
The production of the $J/\psi$ mesons is, however, enhanced by the late regeneration of charmonia, especially at LHC energies.\cite{Chen:2019sak,Abelev:2012rv}
Figure~\ref{fig:OniumFlow} shows the azimuthal anisotropy of $J/\psi$ and, for the first time, of the $\Upsilon$ mesons. 
The $J/\psi$ flow patterns exhibit substantial collective behavior, although less so than D mesons. This is in qualitative agreement with strong charmonium recombination, although it is challenging for models to obtain a quantitative description.\cite{Acharya:2018pjd}
The $\Upsilon(1S)$ state, however, appears to be the only hadron measured at the LHC that shows no $v_2$ within the current precision.\cite{Acharya:2019hlv} This suggests that bottomonia are produced early, and are decoupled from the collectively moving medium, and that late recombination is substantially weaker than in the case of charmonia.

Measurements of baryons containing heavy flavor provide valuable input for theoretical understanding of the heavy-flavor fragmentation. Figure~\ref{fig:LambdaC} presents the charmed baryon to meson ratio $\Lambda_{\rm c}^+/{\rm D}^0$ from recent ALICE measurements in pp collisions. A significant excess is observable in the lower to intermediate $\pT$ range, which cannot be reproduced by commonly used models such as PYTHIA 8\cite{Sjostrand:2014zea}. Note that a similar excess is observed in the $\Xi_{\rm c}^0/{\rm D}^0$ ratio.\cite{Acharya:2017lwf}
As the fragmentation models are tuned using $e^+e-$ collision data, this may raise the question whether heavy-flavor fragmentation is collision system dependent. Some recent model developments, however, are able to capture the observed trends in the $\Lambda_{\rm c}^+/{\rm D}^0$ ratio with either string formation beyond leading color approximation\cite{Christiansen:2015yqa}, or via the feed-down contribution from augmented charmed baryon states\cite{He:2019tik}.

\section{Summary and outlook}

During the Run 1 and Run 2 data taking periods, the ALICE experiment has collected large datasets of pp, p--Pb and nucleus--nucleus collisions at several LHC energies. These data allow for the understanding the system size and energy dependence of hadroproduction, as well as to study the onset of QGP effects and the origin of collective-like behavior in small systems. 
These results contribute to a detailed understanding on the properties of the QGP. This small selection includes intriguing results on global observables that inform us about particle production, bulk property measurements of several species that aim to understand collectivity in the hot matter, and penetrating probes to study energy loss and jet development within the medium. The flavor-dependent studies include precision charm and a wide set of beauty measurements.

After the second long shutdown (LS2), in the Run 3 phase from 2021 on, the LHC will see much improved interaction rate, up to 50 kHz in Pb--Pb collisions. The events will be recorded with upgraded ITS, TPC, MFT and FIT detectors paired with a new continuous readout and computing system. In Run 3 and later in Run 4, altogether an integrated luminosity of 13 nb$^{-1}$ is anticipated by ALICE, which is up to two orders of magnitude more luminosity than Run 1 and Run 2 together. This will allow for a more detailed understanding on the heavy-flavor baryonic sector and a wide range of beauty measurements. With the study of jet structures and event shapes, the soft-hard boundary regime of the strong interaction can be understood in great details.\cite{Noferini:2018are}

\section*{Acknowledgments}

This work has been supported by the Hungarian NKFIH/OTKA K 120660 grant and the J\'anos Bolyai scholarship of the Hungarian Academy of Sciences.

\end{document}